\documentclass[aps,prd,twocolumn,preprintnumbers,superscriptaddress,nofootinbib,floatfix]{revtex4-1}

\usepackage{amsmath}
\usepackage{bm}
\usepackage{amsfonts}
\usepackage{graphicx}
\usepackage{epstopdf}
\usepackage{array}
\usepackage{soul}
\usepackage{color}
\usepackage{hyperref}

\smallskipamount

\enlargethispage{\baselineskip}

\hypersetup{
    colorlinks=true,
    urlcolor=blue,
    citecolor=blue,
    linkcolor=blue,
    menucolor=blue,
    anchorcolor=blue,
    filecolor=blue
}

\everymath{\displaystyle}
\begin{document}


\title{\bf Cavity Detection of Gravitational Waves: Where Do We Stand?}

\newcommand{\TDLI}{\affiliation{Tsung-Dao Lee Institute (TDLI), No.\ 1 Lisuo Road, 201210 Shanghai, China}}
\newcommand{\SJTU}{\affiliation{School of Physics and Astronomy, Shanghai Jiao Tong University, \\ Dongchuan Road 800, 201240 Shanghai, China}}

\author{Claudio Gatti}
\email{claudio.gatti@lnf.infn.it}
\affiliation{INFN Laboratori Nazionali di Frascati, via Enrico Fermi 54, 00044 Frascati (Roma), Italy}

\author{Luca Visinelli}
\email{luca.visinelli@sjtu.edu.cn}
\TDLI \SJTU

\author{Michael Zantedeschi}
\email{zantedeschim@sjtu.edu.cn}
\TDLI \SJTU

\date{\today}

\begin{abstract}
High frequency gravitational waves (HFGWs) are predicted in various exotic scenarios involving both cosmological and astrophysical sources. These elusive signals have recently sparked the interest of a diverse community of researchers, due to the possibility of HFGW detection in the laboratory through graviton-photon conversion in strong magnetic fields. Notable examples include the redesign of the resonant cavities currently under development to detect the cosmic axion. In this work, we derive the sensitivities of some existing and planned resonant cavities to detect a HFGW background. As a concrete scenario, we consider the collective signals that originate from the merging of compact objects, such as two primordial black holes (PBHs) in the asteroid mass window. Our findings improve over existing work by explicitly discussing and quantifying the loss in the experimental reach due to the actual coherence of the source. We elucidate on the approach we adopt in relation with recent literature on the topic. Most notably, we give a recipe for the estimate of the stochastic background that focuses on the presence of the signal in the cavity at all times and showing that, in the relevant PBH mass region, the signal is dominated by coherent binary mergers.
\end{abstract}

\maketitle

\section{Introduction}
\label{sec:introduction}

The detection of gravitational waves (GWs) from compact mergers, made possible via a network of ground-based interferometers, has marked the dawn of GW astronomy~\cite{LIGOScientific:2016aoc, LIGOScientific:2018mvr}. At present, these efforts focus on the sub-kHz GW frequency band, corresponding to the range in reach of the Laser Interferometer Gravitational-Wave Observatory, the Virgo interferometer, and the Kamioka Gravitational Wave Detector.

Nevertheless, the GW spectrum is expected to extend over various decades in frequencies. In fact, the detection in the nHz region of a GW background (GWB) has been recently confirmed by several consortia operating at the global scale, through pulsar timing array techniques~\cite{McLaughlin:2013ira, Goncharov:2021oub, 2016ASPC..502...19L, Chen:2021rqp}. Moreover, the cosmic microwave background (CMB) provides with an indirect constrains on the primordial GWB spectrum at frequencies below the pHz~\cite{BICEP2:2015nss, BICEP2:2018kqh, Namikawa:2019tax}.

Near-future experiments plan to scan different bands encompassing different methods including the forthcoming ground-based~\cite{Punturo:2010zz, Hild:2010id} and space-based~\cite{Yagi:2011wg, LISA:2017pwj} laser interferometers, atom interferometers~\cite{Badurina:2019hst, AEDGE:2019nxb, 2021QS&T....6d4003A}, and probes of the CMB~\cite{Namikawa:2019tax, CMB-S4:2020lpa}. These endeavors cover the GW bands at the kHz frequency and below, where astrophysical and cosmological sources from the merger of known compact objects are expected to provide with a GWB besides other possible sources of unknown origin.

A parallel search can be designed to involve high frequency gravitational waves (HFGWs), spanning frequency ranges well above the kHz. A GWB at high frequencies could potentially be sourced by a collection of exotic physical phenomena originating both in the early and late Universe. Examples include the merging of primordial black holes (PBHs)~\cite{Dolgov:2011cq, Fujita:2014hha, Herman:2020wao, Herman:2022fau, Gehrman:2022imk, Franciolini:2022htd, Barrau:2023kuv, Banerjee:2023brn} or boson stars~\cite{Liebling:2012fv, Visinelli:2021uve}, black hole (BH) superradiance~\cite{Arvanitaki:2010sy}, models of modified gravity~\cite{Jung:2020aem, Lambiase:2022ucu}, the primordial thermal plasma~\cite{Ringwald:2020ist}, phase transitions in the early Universe~\cite{Caprini:2015zlo, Caprini:2018mtu, Addazi:2023jvg}, the ``slingshot'' mechanism, taking place upon the coexistence of confined and unconfined vacua in the presence of heavy quarks~\cite{Bachmaier:2023wzz}, a network of cosmic strings~\cite{Kibble:1976sj, Vilenkin:1981bx, Vilenkin:1984ib, Witten:1984eb, Servant:2023tua}, post-inflation (p)reheating~\cite{Figueroa:2017vfa, Adshead:2018doq}, and inflationary mechanisms~\cite{Kim:2004rp, Vagnozzi:2022qmc}.

A novel avenue for detecting GWs at such high frequencies in the GW spectrum has gained momentum in recent years, with various proposal being explored including interferometers~\cite{Cruise:2006zt, Akutsu:2008qv, Nishizawa:2007tn, Ackley:2020atn}, microwave and optical cavities~\cite{Pegoraro:1978gv, Caves:1979kq, Reece:1984gv, Ballantini:2003nt, Mensky:2009zz, Navarro:2023eii}, mechanical resonators~\cite{Aguiar:2010kn, Goryachev:2014nna, Goryachev:2014yra, Goryachev:2021zzn, 2023NatSR..1310638C, Berlin:2023grv}, superconducting rings~\cite{Anandan:1982is}, and superconducting resonant cavities~\cite{Romanenko:2023irv}. Insightful reviews are found in Refs.~\cite{Aggarwal:2020olq, Schmieden:2023fzn}. 

One promising line that has recently been proposed involves the conversion of gravitons into photons via the inverse Gertsenshtein effect~\cite{Gertsenshtein:1962, Zeldovich:1973, Palessandro:2023tee}. This intriguing concept has positioned resonant cavities as promising candidates for detecting gravitational wave signals within the MHz-GHz HFGW band~\cite{Aggarwal:2020olq}. Previous experiments such as Explorer at CERN, Nautilus at INFN-LNF, and Auriga at Legnaro predominantly targeted gravitational waves in the kHz range, emanating from the mergers of compact objects or comprising a GWB~\cite{Astone:1993ur, Astone:1997gi}. Cavity experiments extend the detection capability to higher frequencies, potentially spanning the range (0.1-10)\,GHz. This groundbreaking approach not only widens the spectrum of detectable gravitational waves but also paves the way for exploring phenomena such as the existence and distribution of PBHs within this yet uncharted frequency domain.

Various cavity searches are currently undergoing, including the Axion Dark Matter eXperiment (ADMX)-G2~\cite{ADMX:2018gho, ADMX:2019uok, ADMX:2021nhd}, ADMX EFR~\cite{Chakrabarty:2023rha}, the Oscillating Resonant Group AxioN Experiment (ORGAN)~\cite{Goryachev:2021zzn, Quiskamp:2022pks, Quiskamp:2023ehr}, the Haloscope At Yale Sensitive To Axion CDM (HAYSTACK)~\cite{Brubaker:2017ohw}, the Center for Axion and Precision Physics Research (CAPP)-8T~\cite{Lee:2020cfj, Choi:2020wyr}, CAPP-9T~\cite{Jeong:2020cwz}, CAPP-PACE~\cite{CAPP:2020utb}, CAPP-18T~\cite{Lee:2022mnc}, CAST-CAPP~\cite{Adair:2022rtw}, GrAHal~\cite{Grenet:2021vbb}, RADES~\cite{Melcon:2018dba, AlvarezMelcon:2020vee, CAST:2020rlf}, TASEH~\cite{TASEH:2022vvu}, QUAX~\cite{Barbieri:2016vwg, Crescini:2018qrz, Alesini:2019ajt, QUAX:2020adt, Alesini:2020vny, Alesini:2022lnp}, and Dark SRF~\cite{Romanenko:2023irv}. Other searches planned to be operational in coming years are the FINUDA magnet for Light Axion SearcH (FLASH)~\cite{Alesini:2019nzq, Alesini:2023qed}, the  International Axion Observatory (IAXO) in its intermediate stage BabyIAXO~\cite{IAXO:2020wwp, Ahyoune:2023gfw}, the Axion Longitudinal Plasma HAloscope (ALPHA)~\cite{Lawson:2019brd, ALPHA:2022rxj}, A Broadband/Resonant Approach to Cosmic Axion Detection with an Amplifying B-field Ring Apparatus (ABRACADABRA)~\cite{Ouellet:2018beu}, DM-Radio~\cite{DMRadio:2022pkf, DMRadio:2022jfv}, the Canfranc Axion Detection Experiment (CADEx)~\cite{Aja:2022csb}, and the magnetized disk and mirror axion experiment (MADMAX)~\cite{Caldwell:2016dcw}. A proposed network of cavities that employs ``quantum squeezing'' would lower the noise and boost the efficiency of the search~\cite{HAYSTAC:2020kwv, Chen:2021bgy, Brady:2022bus, Jiang:2022vpm, Chen:2023ryb}. These endeavors collectively represent a concerted effort to advance our understanding of high-frequency gravitational waves and their potential implications in fundamental physics and cosmology.

In this work, we derive the sensitivities for some of the cavities above. We comment on the detection techniques for either a GW signals or a stochastic background. As a practical example, we consider the GW signal released from BH mergers. This system is characterized by its potentially fast frequency swiping which, contrary to the axion case, can lead to a significant loss of the cavity reach~\cite{Domcke:2022rgu, Franciolini:2022htd}. Our goal is to quantify explicitly such a suppression, detailing the procedures outlined in a previous experimental report~\cite{Alesini:2023qed}.

While our approaches may seem straightforward, we believe this to be worth explicitly commenting on, given the above mentioned advancement and the excitement in the realm of HFGWs. As it turns out, the actual reach given by physical sources such as PBHs mergers is significantly worse than what previously discussed in the literature, see also Refs.~\cite{McNeill:2017uvq, Lasky:2021naa, Domenech:2021odz}. Even in the best case scenario, the discrepancy between current experimental reaches and the physical GW signal amounts to about eight orders of magnitudes, for the case of BH mergers. Rather than interpreting our results as a negative outcome for the detection of HFGWs, we envision them as a catalyst for inspiring novel experimental setups and the study of GW sources.

Our paper is organized as follows. In Sec.~\ref{sec:method} we outline the details of the expression used in the analyses. The results of our approach are presented in Sec.~\ref{sec:results} and a discussion is developed in Sec.~\ref{sec:discussion}. Conclusions are drawn in Sec.~\ref{sec:conclusions}. We set $\hbar = c = 1$ unless otherwise stated.

\section{Method}
\label{sec:method}

\subsection{Sensitivity forecast}

The coupling of the photon with gravity is described by the Maxwell-Einstein action,
\begin{equation}
    \label{eq:MEaction}
    S = \int {\rm d}^4x \sqrt{-g}\left( - \frac{1}{4}g_{\mu\alpha}g_{\nu\beta}F^{\mu\nu}F^{\alpha\beta}\right)\,,
\end{equation}
where $g_{\mu\nu}$ is the space-time metric with determinant $g$ and $F^{\mu\nu}$ is the electromagnetic field strength. 

The term in Eq.~\eqref{eq:MEaction} modifies the usual expressions of electrodynamics by introducing a new source, see Eq.~\eqref{eq:maxwell} in Appendix~\ref{sec:appendixA}, and leads to a deposition of energy in the resonant cavity. To see this, the metric tensor is expanded to first order around a flat background as
\begin{equation}
    g^{\mu\nu} = \eta^{\mu\nu} + h^{\mu\nu}\,,
\end{equation}
where $|h^{\mu\nu}| \ll 1$ describes the perturbations. The action in Eq.~\eqref{eq:MEaction} predicts a coupling between the GW signal and the electromagnetic (EM) energy tensor $T^{\mu\nu}_{\rm EM}$ with the Lagrangian $\mathcal{L} = (1/2)h_{\mu\nu} T^{\mu\nu}_{\rm EM}$. This leads to the effective coupling $\mathcal{L} \propto h_0 B_0\delta B_z$ for a GW strain $h_0$ in an external magnetic field $B_0 \hat{\bf z}$, where the magnetic field variation coupled within the cavity $\delta B_z$ can be picked up with a magnetometer. 

To address the detection of a GW source by a resonating cavity we follow the discussion in Ref.~\cite{Allen:1997ad}, adapting the treatment originally expressed in terms of the search through interferometers to the case of a haloscope. Given the energy stored in the cavity $U$, the signal-to-noise ratio (SNR) is obtained as
\begin{equation}
    \label{eq:SNR}
    {\rm SNR} = \frac{2\pi U}{T_{\rm sys}}\,\sqrt{\Delta t\,\Delta f}\,,
\end{equation}
where $T_{\rm sys}$ is the coldest effective temperature of the instrumentation and noise amplifier, $\Delta f$ is the frequency bandwidth, and $\Delta t$ is the time under which the signal is collected. The expression for the energy stored in the cavity is detailed in Eq.~\eqref{eq:energystored}.

To begin with, we assume that we observe a sufficiently stationary source of GWs for a number of cycles $N_{\rm cycles} \gtrsim 1$. The source emits within the cavity frequency bandwidth $\Delta f = f/Q$, where $Q$ is the quality factor of the cavity, so that the reach of the strain $h_0$ is obtained by setting ${\rm SNR} =1$. Once the expression for the power of the signal in Eq.~\eqref{eq:signal} is considered, this leads to an estimate for the strain at resonance as
\begin{equation}
    \label{eq:strain}
    \begin{split}
    h_0 &\approx 6.0\times10^{-23}\,\left(\frac{\rm 1\,T}{B_0}\right)\,\left(\frac{0.14}{\eta}\right)\,\left(\frac{\rm m^3}{V}\right)^{5/6}\,\left(\frac{10^6}{Q_{\rm eff}}\right)^{1/2}\\
    &\times \left(\frac{T_{\rm sys}}{\rm K}\right)^{1/2}\,\left(\frac{\Delta f}{\rm kHz}\,\frac{\rm 1\,min}{t_{\rm eff}^{\rm int}}\right)^{1/4}\,\left(\frac{\rm GHz}{\omega_n/2\pi}\right)^{3/2}\,,
    \end{split}
\end{equation}
where $Q_{\rm eff}= {\rm min}(N_{\rm cycles},Q)$ is the effective quality factor, $\eta$ the coupling of the cavity with the gravitational signal defined in Eq.~\eqref{eq:defineeta}, $V$ the effective volume of the cavity, $T_{\rm sys}$ the system temperature, $\omega_n= 2\pi f$ the resonant pulsation, and $t_{\rm eff}^{\rm int}= {\rm min}(N_{\rm cycles}\,\omega_n^{-1},\Delta t)$ the minimum value between the experimental integration time and the source duration. For a detailed derivation and definition of all the above quantities we refer the reader to Appendix~\ref{sec:appendixA}.
The expression in Eq.~\eqref{eq:SNR} can also be used to estimate the sensitivity over a stochastic GW background. A GWB spectrum is generally assumed to be nearly isotropic, unpolarized, stationary, and characterized by a Gaussian distribution with zero mean. The fractional contribution to the density parameter $\Omega_{\rm GW}$ can be alternatively expressed in terms of a dimensionless characteristic strain $h_c$ as~\cite{Thrane:2013oya, Romano:2016dpx}
\begin{equation}
    \label{eq:OmegaGW}
    \Omega_{\rm GW}(f) = \frac{(2\pi)^2}{3H_0^2}\,f^2\,h_c^2\,,
\end{equation}
where $H_0$ is the Hubble constant.

As we discuss in Appendix~\ref{sec:appendixA}, provided $Q_{\rm eff}\sim Q$ - whose validity depends on the properties of the sources discussed below - the characteristic strain reach reads
\begin{equation}
    \label{eq:strain_hc}
    \begin{split}
    h_c &\approx 6.0\times10^{-20}\,\left(\frac{\rm 1\,T}{B_0}\right)\,\left(\frac{0.14}{\eta}\right)\,\left(\frac{\rm m^3}{V}\right)^{5/6}\\
    &\times \left(\frac{T_{\rm sys}}{\rm K}\right)^{1/2}\,\left(\frac{\Delta f}{\rm kHz}\,\frac{\rm 1\,min}{t_{\rm obs}}\right)^{1/4}\,\left(\frac{\rm GHz}{\omega_n/2\pi}\right)^{3/2},
    \end{split}
\end{equation}
where the total observation time of the experiment $t_{\rm obs}$ can strategically largely exceed the value chosen for the search for a coherent source.

Note, that the studies in Refs.~\cite{Herman:2020wao, Herman:2022fau} find a much smaller sensitivity for the GW strain $h\sim 10^{-30}$, which would be ideal to probe the floor of GW background from coalescing compact objects of asteroidal mass. It is unclear to us how their analysis in time-domain can increase the experimental reach in such a non-trivial manner, given the analysis leading to Eq.~\eqref{eq:strain_hc}. However, as stated in their conclusions, several assumptions are made regarding the coherence and duration of the stochastic background. Furthermore, the problem of frequency width of a stochastic GW background, as compared to the narrow width of the cavity, remains unaddressed. In this work, we characterize some of these effects.

So far, we have discussed the response of the cavity to a stationary signal. We now briefly comment on the coherence of the source. For the dark matter axion, the quality of the source is limited by thermal and quantum fluctuations which impact over the maximal capacity of the cavity to resonate with the signal. For the case of a GW signal from PBH mergers or other sources of HFGWs, the actual coherence of the GW signal at a given frequency is granted by the large number of gravitons in the parameter space of interest. This implies that the same approach as for the axion can be used to parametrize the non-stationarity of a HFGW source.

In fact, we can describe the gravitational wave signal as a coherent state of highly occupied gravitons of energy $\omega$, with the occupation number $N_g\sim \rho/\omega^4\sim m_{\rm Pl}^2 h_0^2\,/\omega^2$ corresponding to the number of gravitons per de Broglie volume $\omega^{-3}$. Here, we have used the fact that the energy density of the GW signal is given by $\rho \sim h_0^2 \omega^2 m_{\rm Pl}^2$. For a coherent signal, typical fluctuations are of the order of $\delta N_g \sim \sqrt{N_g}$, leading to a quality of the signal $Q_h = N_g/ \delta N_g \sim \sqrt{N_g}$. This quality factor is potentially degrading the quality of the source when it is smaller than the quality of the cavity $Q$. Therefore, the quality of the source is determined by the quantum origin of the signal.

For example, consider a GW signal with the frequency $f \sim 0.1$\,GHz and a typical physical strain of the reach $h_0 \sim 10^{-22}$, corresponding to the typical current reach of the cavity given in Eq.~\eqref{eq:strainth} below. In this setup, the number of gravitons is $N_{\rm g} \sim 10^{32}$, which implies a source quality $Q_h\sim 10^{16}$ that is much larger than the quality factor of the cavity. Even in the desirable scenario where the cavity would reach a sensitivity comparable to the actual physical signal of strain $h_0\sim 10^{-30}$, the source would possess the quality $Q_h \sim 10^6 \gtrsim Q$. The consideration above therefore justifies the dropping of such a contribution from the treatment. Note, however, that if the sensitivity of the experiment could reach an even smaller sensitivity for this range of frequencies, part of the signal could be degraded.

\subsection{Gravitational wave sources}

Potential sources of GWs are generally divided into two categories, namely sources of cosmological origin produced before recombination and sources of astrophysical origin. A cosmological GWB at high frequencies, expected from exotic sources, can be constrained by BBN considerations and CMB data as~\cite{Caprini:2018mtu}
\begin{equation}
    \label{eq:BBNbound}
    \Omega_{\rm GW}h_H^2 \lesssim 5.6\times 10^{-6}\,\Delta N_{\rm eff}\,,
\end{equation}
where $h_H = H_0/(100{\rm\,km\,s^{-1}\,Mpc^{-1}})$ is the reduced Hubble constant. The excess in the number of relativistic active neutrinos is constrained as $\Delta N_{\rm eff} \lesssim 0.3$ by various considerations on big-bang nucleosynthesis (BBN) in combinations with CMB data. Using Eq.~\eqref{eq:OmegaGW}, the bound above reads~\cite{Domcke:2022rgu, Domcke:2023bat}
\begin{equation}
    h_c \lesssim 2\times 10^{-30}\,\left({\rm GHz}/f\right)\,\Delta N_{\rm eff}^{1/2}\,,
\end{equation}
which is several orders of magnitude below the expected reach of resonant cavities expressed in Eq.~\eqref{eq:strain_hc}. Therefore, a potential detection of HFGWs can only be due to astrophysical sources in the late Universe.

However, since there are no known astrophysical sources releasing GWs at such high frequencies, new physics is likely required to motivate the search in the HFGW band. Possibilities include the decay of an unstable axion star after a binary merging~\cite{Chung-Jukko:2024hod} and the stimulated decay of dark matter (DM) in theories of Chern-Simons gravity~\cite{Jung:2020aem}. We comment below on the merging of compact objects, with particular focus on the case of PBHs. We consider BHs that are too light to be explained by known stellar dynamics, so that their indirect detection through mergers would require new physics to explain their origins. A possibility is that these PBHs form in the early Universe, hence the name primordial. The exact details of the formation scenario are currently unknown. For example, they could result from inflationary overdensities~\cite{Carr:2020gox}, from the collapse of bubbles in supercooled phase transitions~\cite{Gouttenoire:2023naa,Flores:2024lng}, from the confinement of heavy quarks~\cite{Dvali:2021byy}, or many other methods~\cite{Carr:2020gox}.

In Table~\ref{tab:sensitivity} we report the main setups of some experiments that employ a resonant cavity, see the caption for the specific references. For each experiment, the noise temperature $T_{\rm sys}$ is obtained through the formula $T_{\rm sys} = \omega_c[1/(\exp(\omega_c/T_{\rm phys}) -1)+N_A+0.5]$, where $\omega_c/(2\pi)$ is the central frequency in the band covered by the experiment and the number of states is $N_A = 1/2$ everywhere except for FLASH and BabyIAXO, for which $N_A = 10$. In this context, ``FLASH HighT'' refers to the initial phase of the experiment~\cite{Alesini:2023qed}. This phase will undergo an upgrade with an enhanced cryostat to achieve the performance levels of ``FLASH LowT'', which serve as benchmarks for the results discussed below.\footnote{A parallel search at the higher frequency range (206–360)\,MHz with a lower volume of $1.312{\rm\,m^3}$ is also planned.}
\begin{table*}[!ht]
    \def\arraystretch{1.2}
    \begin{center}
    \vspace*{0.5cm}
    \begin{tabular}{c|c c c c c c}
      Experiment & Frequency range [GHz] & Volume [m$^3$] & Unloaded $Q$/10$^3$ & $B_0$\, [T] & $T_{\rm phys}$\, [K] & $T_{\rm sys}$\, [K]\\
      \hline
      ADMX-G2 & 0.6–2 & 0.085 & 60 & 7.7 & 0.15 & 0.18\\
      ADMX EFR & 2-4 & 0.080 & 90 & 9.6 & 0.15 & 0.23\\
      FLASH HighT & 0.117-0.206 & 4.156 & 570–450 & 1.1 & 4.9 & 5.0\\
      FLASH LowT & 0.117-0.206 & 4.156 & 570–450 & 1.1 & 0.30 & 0.40\\
      ALPHA & 5-40 & 0.334 & 10 & 13 & 5.0 & 5.0\\
      HAYSTACK & 4-6 & 0.0015 & 30 & 9 & 0.13 & 0.28\\
      BabyIAXO & 0.253-0.469 & 2.7 & 320 & 2 & 4.9 & 4.9\\
      \hline\hline
    \end{tabular}
    \end{center}
    \caption{Parameters defining the resonant cavity experiments considered in this work. Specifically, ADMX-G2~\cite{ADMX:2018gho, ADMX:2019uok, ADMX:2021nhd}, ADMX EFR~\cite{ADMXEFR}, FLASH low-frequency (LF) and high-frequency (HF)~\cite{Alesini:2023qed}, ALPHA~\cite{Lawson:2019brd, ALPHA:2022rxj}, HAYSTACK~\cite{Brubaker:2017ohw}, and BabyIAXO~\cite{IAXO:2020wwp, Ahyoune:2023gfw}.}
    \label{tab:sensitivity}
\end{table*}

A fundamental question arises regarding whether these objects, regardless of how they formed, can constitute a substantial portion of DM. Cavity searches are sensitive to mergers involving sub-solar PBHs with masses ranging below about $10^{-8}\,M_{\odot}$. This result, as we outline below, derives from requiring that at least one complete revolution of the binary system appears in the cavity tuned at the frequency around the GHz. This PBH mass range partially overlaps with the region of masses heavier than $10^{-11}\,M_\odot$, in which stringent lensing constraints have already discounted PBHs as the primary constituents of DM~\cite{Niikura:2017zjd}. To sum up, in the region of masses $[10^{-11}\textrm{-}10^{-8}]\,M_\odot$ a minor contribution from PBHs to the DM abundance is still plausible~\cite{Carr:2020gox}, while the region of masses below $10^{-11}\,M_{\odot}$ is not currently constrained by microlensing results.

Although cavity experiments do not reach the sensitivity required to probe the type of signal from these sources, several uncertainties in cosmological history could substantially amplify the signal. One explicit possibility under discussion is the incorporation of significant non-Gaussianity from the inflationary period, which, in turn, could lead to an escalation in the merger rate~\cite{Franciolini:2022htd}. Note, that even in the best scenario a maximal increase of about two orders of magnitude is feasible in the merger rate~\cite{Franciolini:2022htd}.

Compact objects such as BH and neutron stars (NS) form in astrophysical environment. Possible and more exotic configurations such as PBHs or boson stars could have already been present in the earliest stages of the Universe~\cite{Liebling:2012fv, Visinelli:2021uve}. Binaries of compact objects could fall in the frequency range and with a GW strain that is sufficiently strong to be detectable with present or near-future technologies. Consider two compact objects of similar mass $M$ and size $R$, and each of compactness $\mathrm{C} \equiv GM/R$, forming a system of total mass $M_{\rm TOT} \approx 2M$. The frequency of the emitted GW spectrum at the end of the inspiral phase, when the stars occupy the innermost stable circular orbit, is~\cite{Giudice:2016zpa}
\begin{equation}
	\label{eq:fISCO}
	f = \frac{\mathrm{C}^{3/2}}{3\sqrt{3}\pi G M_{\rm TOT}}\,.
\end{equation}
For example, a signal in the bandwidth $\mathcal{O}(100\,{\rm MHz})$ gives
\begin{equation}
\label{eq:mtotexample}
    M_{\rm TOT} \approx 4\times 10^{-6}\,M_\odot\,\left(\frac{\mathrm{C}}{0.1}\right)^{3/2}\,,
\end{equation}
which, for the compactness of a BH $\mathrm{C}_{\rm BH} = 0.5$, corresponds to the frequency of PBH binaries with mass $M_{\rm BH} \sim 10^{-5}\,M_{\odot}$.

\section{Results}
\label{sec:results}

We focus on the detection of light PBHs in the asteroid mass window, whose merging would lead to the release of a HFGW signal with the GW strain~\cite{Franciolini:2022htd}
\begin{equation}
    \label{eq:strainth}
    h_0 \approx 10^{-22}\left(\frac{10\,\rm kpc}{d}\right)\!\left(\frac{M_{\rm TOT}}{2.2\times10^{-5}\,M_\odot}\right)^{\frac{5}{3}}\!\left(\frac{f}{200\,\rm MHz}\right)^{\frac{2}{3}}.
\end{equation}
The distance $d$ is fixed by requiring that at least one PBH merger per year occurs in the Galaxy and assuming $f_{\rm PBH} \equiv \rho_{\rm PBH}/\rho_{\rm DM}=1$, where $\rho_{\rm DM}$ is the DM density today. In the derivation of the merger rate, a further enhancement coming from the galactic overdensity has been accounted for~\cite{Pujolas:2021yaw}. See also Fig.~3 in Ref.~\cite{Franciolini:2022htd}.

An event within a distance $d\lesssim 1$\,kpc could therefore be within the reach of the cavity experiments. Unfortunately, the source of GW, namely the in-spiraling binary, cannot be treated as a coherent source. The system emits at a given frequency for a number of cycles given by~\cite{Moore:2014lga}
\begin{equation}
\label{eq:Ncycles}
    {N_{\rm cycles}^{\textrm{na{\"i}ve}}} = \frac{f^2}{\dot{f}}\simeq  \left(\frac{M}{2.2 \times 10^{-5}\,M_{\odot}} \right)^{-5/3}\left(\frac{f}{200\,\rm MHz} \right)^{-5/3}.
\end{equation}
This can result in an effective limitation of the source at resonating with the detector. Eq.~\eqref{eq:Ncycles} describes the number of oscillations performed by the source at a given frequency, in a frequency width of order $f$ during the swiping. In the literature, this is assumed to be the number of cycles for which the merger is a proper source inside of the cavity~\cite{Franciolini:2022htd}. However, a merger can only resonate in a cavity as long as its frequency lies within the frequency width of the cavity $\Delta f\ll f$ itself. This leads to a further loss of reach, as the effective number of cycles within the cavity is expressed as
\begin{equation}
    \label{eq:Ncycleseff}
    N_{\rm cycles} = \frac{\Delta f}{f}{N_{\rm cycles}^{\textrm{na{\"i}ve}}}\simeq  \frac{1}{Q}{N_{\rm cycles}^{\textrm{na{\"i}ve}}}.
\end{equation}
As we discuss below, the expression in Eq.~\eqref{eq:Ncycleseff} forces the reach of the detector described in Eq.~\eqref{eq:strain} to be valid upon replacing the quality factor with the effective quantity $Q_{\rm eff} \equiv \min(Q,N_{\rm cycles})$. Similar comments regarding the swiping time $t_{\rm int}^{\rm eff}$ and the value of $Q_{\rm eff}$ have been pointed out in Ref.~\cite{Barrau:2023kuv}. Our work extends the discussion by pointing out that the optimal mass range for the detection of GWs from PBH mergers relies on maximizing the cavity resonance.

\begin{figure*}[th!]
    \centering
    \includegraphics[width=\textwidth]{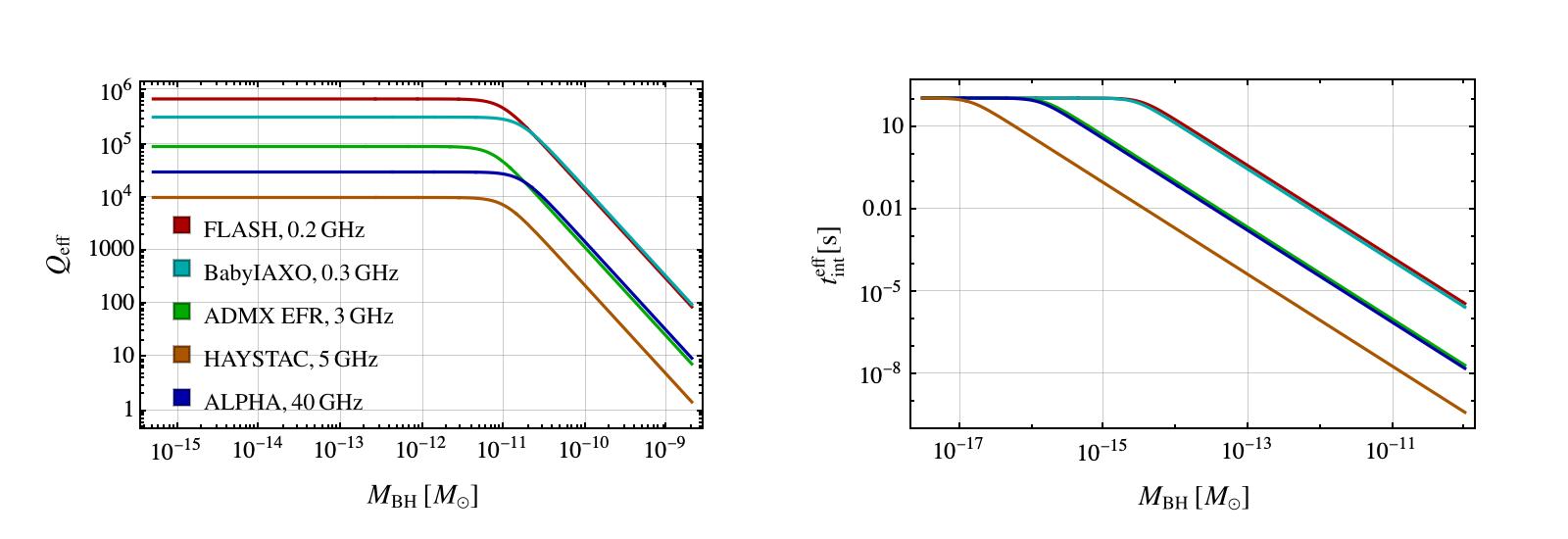}
    \caption{The effective quality factor $Q_{\rm eff}$ (left panel) and the minimum value between the source duration and the experimental observation time $t_{\rm obs}$, here $t_{\rm int}^{\rm eff}$ (right panel) for the experiments listed in Table~\ref{tab:sensitivity}. As a reference, the value $t_{\rm obs} = 120$\,s is chosen.}
    \label{fig:suppressionqeff}
\end{figure*}

Fig.~\ref{fig:suppressionqeff} shows the effective quality factor $Q_{\rm eff}$ (left panel) as a function of the PBH binary merger mass, at fixed frequencies for the experiments in Table~\ref{tab:sensitivity}.\footnote{The mass of a PBH is bound from below by considerations over its evaporation byproducts, see e.g.\ Ref.~\cite{Coogan:2020tuf}.} A maximal resonance is possible in this class of experiments only for PBH mergers lighter than about $10^{-11}\,M_{\odot}$. For heavier mergers, the number of cycles scales as $M^{-5/3}$, leading to the complete absence of resonance $Q_{\rm eff}\sim 1$ for masses around $10^{-9}\,M_{\odot}$. No detection is possible for heavier PBH mergers with current strategies, perhaps suggesting the adoption of broadband type of experiments in those region, as discussed e.g.\ in Ref.~\cite{Domcke:2022rgu,Franciolini:2022htd,Tobar:2022pie}.

Therefore, two competing effects come into play when the detection of PBH mergers is considered. On one hand, at a fixed frequency one would like to consider heavier binaries, as the signal expected in Eq.~\eqref{eq:strainth} would be more prominent for at higher masses. On the other hand, it would be optimal to exploit the cavity resonance with the highest possible $Q_{\rm eff}$, thus requiring the search for light PBH binaries. As discussed below, it is indeed the former effect determining the optimal reach to be around $10^{-11}\,M_{\odot}$ given the physical signal of PBH mergers.

Another factor impacting the experimental reach for an individual source is expressed by the effective integration time $t_{\rm int}^{\rm eff}$, which is here defined as the minimum value between the source duration and the experimental observation time $t_{\rm obs}$. Here, we set $t_{\rm obs} = 120$\,s as a reference. The effective integration time is typically of the order of a few minutes at a given frequency and is limited by the swiping time of the PBH merger within the frequency width of the cavity. This is shown in the right panel of Fig.~\ref{fig:suppressionqeff}. The effect is in place for basically all PBH mergers with a total mass heavier than $10^{-14}\,M_{\odot}$. A direct inspection of Eq.~\eqref{eq:strainth} shows that the effect is not as impactful as the decrease in sensitivity coming from the scaling of $Q_{\rm eff}$ with the binary mass.

Compact objects that come close to each other in unbound orbits perform a hyperbolic motion, leading to a single scattering event that manifests itself through a burst of GWs~\cite{Garcia-Bellido:2017qal, Garcia-Bellido:2021jlq, Teuscher:2024xft, Barrau:2024kcb}. The potential reach due to hyperbolic encounter does not greatly differ from the case of bound orbits, since the duration of the signal resonating in the cavity is extremely short. When considered, this additional contribution would affect the detectable strain by a factor of the order of $\mathcal{O}(1)$, so that the conclusions drawn above would not be altered considerably.

A significant enhancement of Eq.~\eqref{eq:strainth} would be brought in by the possible presence of a non-Gaussianity component in the density perturbations~\cite{Franciolini:2022htd}. In the region of optimal reach around $10^{-11}\,M_{\odot}$, the maximal enhancement is similar in amplitude, to the decrease in signal due to the typical bounds on lensing, requiring $f_{\rm PBH}\sim \mathcal{O}(10^{-2}-10^{-3})$ in the PBH population. This justifies the adoption of $f_{\rm PBH}=1$ as an optimistic case scenario.

\section{Discussion}
\label{sec:discussion}

\subsection{Coherent sources}

\begin{figure}[!ht]
    \centering
    \includegraphics[width=\linewidth]{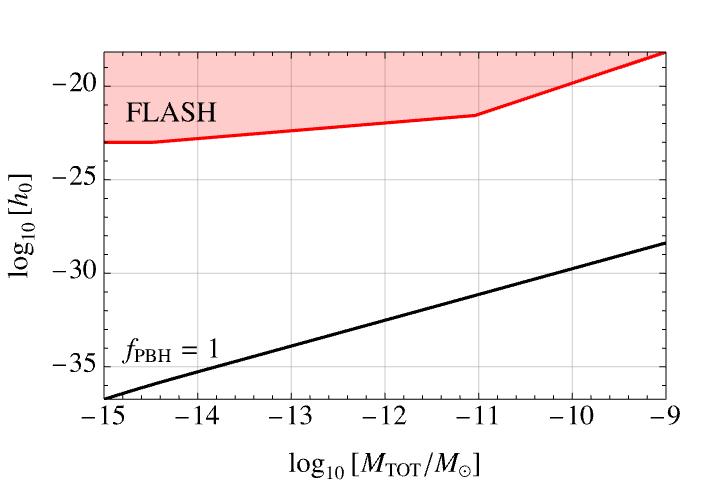}
    \caption{Comparison of the forecast reach for FLASH for different PBH masses. The red line correspond to the reach according to the values in Table~\ref{tab:sensitivity}. The black line corresponds to the theoretical prediction range for a monochromatic PBH distribution of values $f_{\rm PBH}=1$. The cavity frequency is fixed at $f_c = 200$\,MHz.}
    \label{fig:FLASHreach}
\end{figure}

We first discuss the phenomenology related to the detection of a coherent source such as a binary merger. To showcase the effects affecting the cavity reach due to the non-stationary behaviour of the source, we first consider the setup of the resonant cavity employed in the FLASH experiment,\footnote{The FINUDA magnet, a core element for the FLASH experiment, has been recently successfully tested.} as considered in Table~\ref{tab:sensitivity}. The reach for the experiment according to Eq.~\eqref{eq:strain} is shown in Fig.~\ref{fig:FLASHreach} (red solid line) in comparison with the signal generated by mergers with $f_{\rm PBH}=1$ (black solid line). Note, that the expected signal scales linearly with $f_{\rm PBH}$. In Fig.~\ref{fig:FLASHreach}, the frequency is fixed at the cavity value $f_c = 200\,\rm MHz$, which is within reach for the FLASH setup. The drop in reach for $M\lesssim 10^{-11}\,M_\odot$ is due to the non-stationary behaviour of the source, resulting in a lower effective quality factor $Q_{\rm eff}$ as discussed below Eq.~\eqref{eq:Ncycles} and explicitly shown in Fig.~\ref{fig:suppressionqeff}.

Another loss in sensitivity takes place for PBHs lighter than about $10^{-15}\,M_{\odot}$. In this mass region, the limit factor originates from the effective time required by the source to swipe over a frequency width of the order of $f/Q=\Delta f$, as shown in the right panel of Fig.~\ref{fig:suppressionqeff}. In fact, the optimal reach region balancing between the physical signal and the cavity reach is around the $M_{\rm BH} \approx 10^{-11}\,M_{\odot}$ mass window, for which the optimal reach is achieved when
\begin{equation}
    \label{eq:optimalqeff}
    Q_{\rm eff} \simeq N_{\rm cycle}\simeq Q\,.
\end{equation}
Similar results as in Fig.~\ref{fig:FLASHreach} qualitatively hold true for the other cavity setups expressed in Table~\ref{tab:sensitivity}.

\begin{figure*}[th!]
    \centering
    \includegraphics[width=\textwidth]{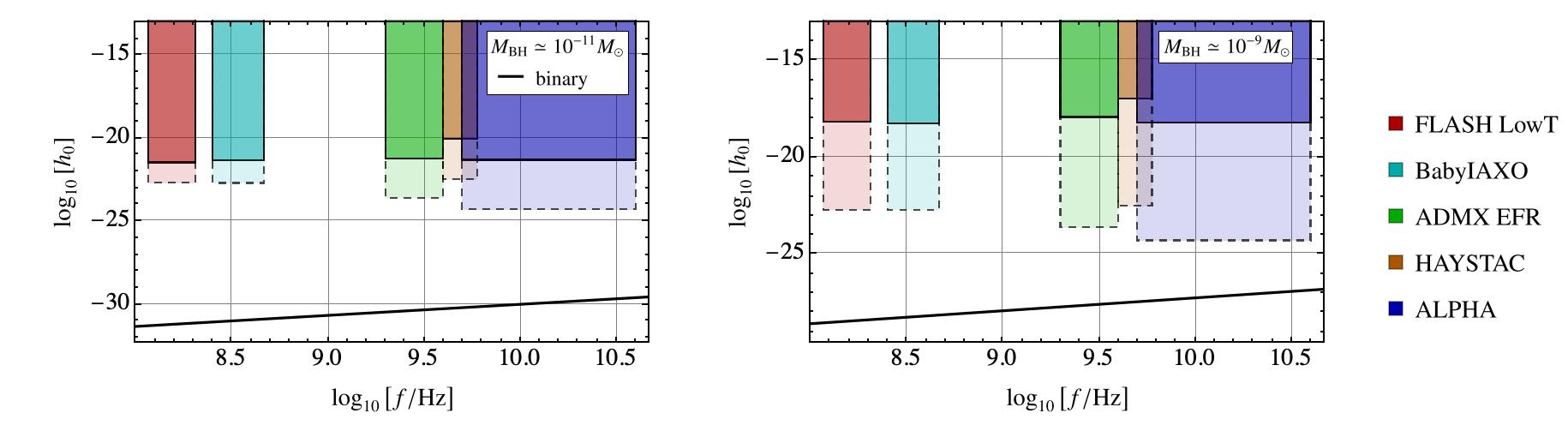}
    \caption{Cavity reach for the experiments listed in Table~\ref{tab:sensitivity} as a function of frequency. For illustration purposes, we report the reach for the central value of the frequency range of each experiment. The lightly-shaded area denotes the expected reach for an ideal source, not limiting the capability of the cavity to fully resonate. These results are consistent with the sensitivity reach found, e.g., in Refs.~\cite{Berlin:2021txa,Domcke:2022rgu,Franciolini:2022htd} which, indeed, relied on the same assumption.} The shaded area denotes the actual reach as compared to the actual source of a binary with mass $10^{-11}\,M_{\odot}$ (left panel) and $10^{-9}\,M_{\odot}$ (right panel), whose physical expected amplitude is marked by a black solid line. The PBH population is assumed to satisfy $f_{\rm PBH}=1$.
    \label{fig:reachhcmergercoherent}
\end{figure*}

We now consider the forecast reach in the cavity experiments given in Table~\ref{tab:sensitivity}. Results for the reach as a function of the cavity frequency are shown in Fig.~\ref{fig:reachhcmergercoherent} for the PBH mass $M_{\rm BH} = 10^{-11}\,M_\odot$ (left) and $M_{\rm BH} = 10^{-9}\,M_\odot$ (right). For each panel, the expected reach (dark shaded area) is compared with the ideal reach of a perfectly coherent source (lightly shaded area). For the binary masses considered, most of the actual reach loss accounts for the frequency swiping time being shorter than the typical integration time. In fact, according to Eq.~\eqref{eq:optimalqeff} the effective quality factor $Q_{\rm eff}$ is maximized in this mass region. This corresponds to an ideal source that allows for the cavity to fully resonate without any loss due to a quality factor at source. This is the reach usually showed in the literature, see e.g.\ Refs.~\cite{Berlin:2021txa, Franciolini:2022htd,Tobar:2022pie}. For each panel, the black solid line marks the physical signal expected by a population of PBH binary mergers obtained using the results in Ref.~\cite{Franciolini:2022htd}. Although cavity experiments swiping higher frequency ranges have a higher chance of working in the regime closer to the actual potential physical signal, there exists a discrepancy by many orders of magnitude from a potential detection of this source.

The right panel in Fig.~\ref{fig:reachhcmergercoherent} shows the dramatic improvement in considering a heavier mass range $M_{\rm BH} \approx 10^{-9}\,M_{\odot}$. While the physical signal increases in strain, the actual reach (dark shaded areas) significantly decreases. This is due to the quality of the source $N_{\rm eff}$ being placed away from the condition in Eq.~\eqref{eq:optimalqeff} for the mass range considered. In this scenario, the physical cavity reach effectively places even further away from the physical signal when compared with the optimal case scenario $M_{\rm BH} \approx 10^{-11}\,M_{\odot}$ previously discussed and shown in the left panel.

\subsection{Stochastic source}

We now turn to the discussion over the stochastic signal sourced by merging PBH binaries, characterized by a superposition of weak, incoherent, and unresolved GW sources~\cite{1987MNRAS.227..933M, Christensen:1992wi, Flanagan:1993ix}. For this signal, the loss in coherence described in the previous section does not affect the reach for the mass region considered. However, the typical strain signal turns out to be significantly lower even for an optimistic scenario where the strain $h_0\lesssim 10^{-26}$ is expected~\cite{Franciolini:2022htd}.

The analysis of a stochastic signal demands the coincident detection by two correlated and co-aligned GW detectors, each picking up a GW strain $h_i(f)$ with $i=1,2$ labeling the detector.\footnote{Building a network of HFGW detectors is required to pick up transient GW events that appear as a coincident detection, distinguishing them from a noise transient.} The signal-to-noise ratio can be expressed as~\cite{Allen:1997ad}
\begin{equation}
    \label{eq:snrth}
    {\rm SNR} \approx \frac{3H_0^2}{10\pi^2}\,\sqrt{t_{\rm obs}}\, \left(\int_{-\infty}^{+\infty} {\rm d}f\,\frac{\gamma^2(f)\,\Omega_{\rm GW}^2(f)}{f^6 \,P_1(f)\,P_2(f)}\right)^{1/2}\,.
\end{equation}
Here, $P_i(f)$ is the noise power spectrum of the $i$-th detector, which is related to the variance of the cross-correlation signal as
\begin{equation}
    \sigma_i^2 = \int_{0}^{+\infty}{\rm d}f\,P_i(f)\,.
\end{equation}
The quantity $\gamma(f)$, known as the overlap reduction function, quantifies the reduction in sensitivity due to the time delay between the two detectors and the non-parallel alignment of the cavity axes~\cite{Flanagan:1993ix}. For coincident and co-aligned detectors it is $\gamma(f) = 1$, while it is expected $\gamma(f) < 1$ when the detectors are shifted apart or rotated.

To claim the successful detection of a GWB, we assume that such a signal is indeed present at the frequencies considered, with a mean value that allows for its correct identification for a fraction $\delta$ of the times. In this context, the SNR should satisfy~\cite{Allen:1997ad}
\begin{equation}
    {\rm SNR} \geq [{\rm erfc}^{-1}(2 \alpha) - {\rm erfc}^{-1}(2 \delta)]/\sqrt{2}\,\,,
\end{equation}
where $\alpha$ quantifies the false alarm rate. Setting a false alarm rate $\alpha = 0.05$ and a detection rate $\delta = 0.95$ leads to the requirement SNR $\gtrsim$ 1.64. Note, that the presence of $t_{\rm obs}$ appearing under the square root in Eq.~\eqref{eq:snrth} allows to consider much longer integration times when searching for the stochastic GWB compared with the coherent signal. Moreover, increasing $t_{\rm obs}$ in a search of a coherent, but transient, signal would not lead to an improved sensitivity because of the shortness of the signal duration.

\begin{figure}[ht!]
    \centering
    \includegraphics[width=\linewidth]{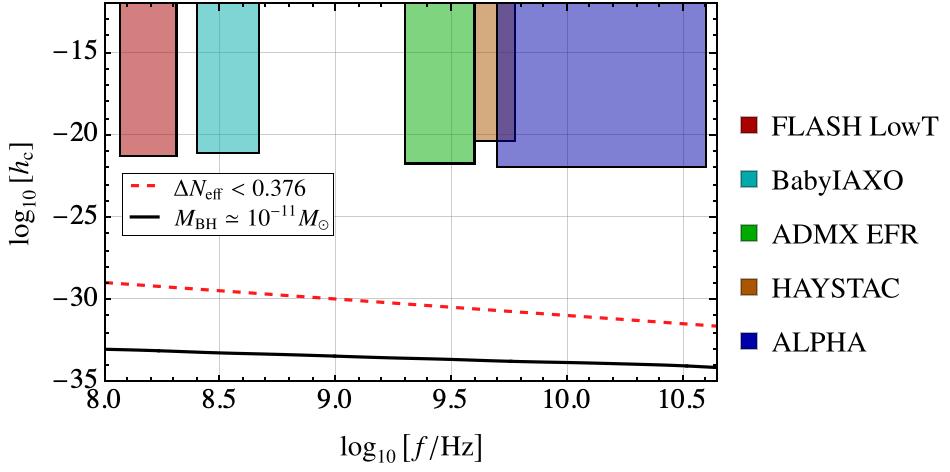}
    \caption{Stochastic GW background reach. The black line denotes the actual physical signal for $f_{\rm PBH}=1$ and the central value of the log-normal distribution $M_c= 10^{-11}\,M_{\odot}$, see Eq.~\eqref{eq:lognormal}. The BBN bound is reported for comparison (dashed red line).}
    \label{fig:omegagwvsreachstochastic}
\end{figure}

We assess the sensitivity of the strain $h_c$ in Eq.~\eqref{eq:strain_hc} assuming a period of observation $t_{\rm obs} = 6\,$months, leading to the results in Fig.~\ref{fig:omegagwvsreachstochastic} for the different experiments in Table~\ref{tab:sensitivity}. To estimate the expected physical signal we adopt the derivation of Refs.~\cite{Ajith:2009bn,Zhu:2011bd}, see also Ref.~\cite{Franciolini:2022htd} for a detailed discussion. We assume that the PBH masses are distributed according to a log-normal distribution centered at $M_{c}= 10^{-11}\,M_{\odot}$ and of width $\sigma= 0.26$ as~\cite{Dolgov:1992pu, Carr:2017jsz}
\begin{equation}
    \label{eq:lognormal}
    \psi(M_{\rm BH})= \frac{f_{\rm PBH}}{\sqrt{2\pi} \sigma M_{\rm BH}}\exp\left[\frac{-\log\left(M_{\rm BH}/M_c\right)^2}{2\sigma^2}\right]\,.
\end{equation}
Note, that an optimal reach is guaranteed for the range of masses considered since $Q_{\rm eff}\sim Q$. The distribution in Eq.~\eqref{eq:lognormal} is normalized so that the fraction of PBHs that make up the DM is
\begin{equation}
    f_{\rm PBH} = \int {\rm d}M_{\rm BH}\,\psi(M_{\rm BH})\,.
\end{equation}
The result for the expected signal shown in Fig.~\ref{fig:omegagwvsreachstochastic} (black solid line) scales linearly with the PBH abundance $f_{\rm PBH}$, which is here set as $f_{\rm PBH} = 1$. Even in this optimistic scenario, the values lie several orders of magnitude below the bound from BBN considerations in Eq.~\eqref{eq:BBNbound} (red-dashed line) and the actual cavity sensitivity of the various experiments.

\subsection{Can we expect a continuous signal?}
\label{sec:continuous}

To address the stochastic GWB from PBH mergers, a monochromatic mass distribution with $f_{\rm PBH}=1$ is now assumed for simplicity, in analogy with the treatment for a coherent signal. Our findings thus allow for a direct comparison between the coherent and the stochastic cases. We generally obtain that for large PBH masses it is more convenient to search for a coherent signal, while the stochastic signal is stronger for light PBH mergers. Note, that for this type of analysis it is required to coordinate at least two detectors~\cite{Aggarwal:2020olq}.

A necessary requirement for a stochastic noise to be detectable is that, at any time, at least $\mathcal{O}(1)$ mergers are sourcing within the cavity. As discussed above, the swiping time $t_{\rm swiping}$ across frequencies of the order of the frequency width of the cavity $\Delta f$ depends on the masses of the BH pair, with the typical $t_{\rm swiping}$ as given in Fig.~\ref{fig:suppressionqeff}. Therefore, at least one merger per swiping time should take place at every moment. This requirement provides the typical distance $d_{\rm swiping}$ at which a merger takes place, therefore allowing for an optimistic characterisation of the amplitude of the signal. 

Given a rate per unit volume of PBH mergers $R_{\rm PBH}$, see e.g.\ Ref.~\cite{Franciolini:2022htd}, the typical distances at which one merger event is constantly present in the cavity is found from integrating over a volume of radius $d_{\rm swiping}= (3 V_{\rm swiping}/4\pi)^{1/3}$, to obtain
\begin{equation}
    \label{eq:stochcondition}
	1 \lesssim t_{\rm swiping} \int_0^{d_{\rm swiping}} {\rm d}r\, 4\pi r^2 R_{\rm PBH}.
\end{equation}
While the above condition might not be sufficient to provide a stochastic signal, it surely suffices. More pragmatically, we address the question whether a continuous signal offers better detection prospects rather than fewer and rarer events of larger amplitude. As a byproduct, we determine the best search strategy for detecting PBH mergers in the mass window considered.

In previous literature the distance has been fixed through a different requirement, more precisely by looking for the volume within which one coherent merger per year is realized~\cite{Franciolini:2022htd}. Under the condition in Eq.~\eqref{eq:stochcondition}, the typical distance of a merger found by our requirement to realize at least one merger at each moment of time is significantly larger than the case of one coherent merger per year. 

\begin{figure}[!ht]
    \centering
    \includegraphics[width=\linewidth]{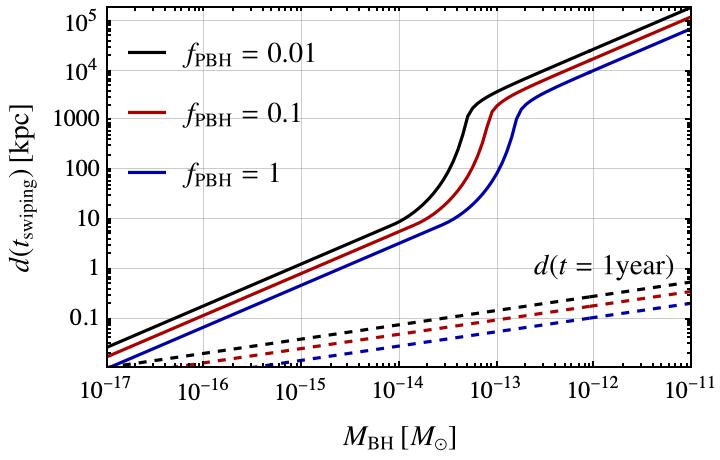}
    \caption{Distance for mergers sourcing a gravitational wave stochastic background under the condition in Eq.~\eqref{eq:stochcondition} (solid line). The coherent case of one merger here is shown with dashed lines.}
    \label{fig:distancemergers}
\end{figure}

In Fig.~\ref{fig:distancemergers}, we show the comparison between the typical distances of black hole mergers sourcing a stochastic background (continuous line) with the coherent case of one merger per year $f_{\rm PBH}$ (dashed line) for different values of $f_{\rm PBH}$. In the latter case, the resulting curve is found to be consistent with the one in Ref.~\cite{Franciolini:2022htd}. In the former case, the distance now also depends on the typical swiping time. For PBH masses above $\sim 10^{-14}M_{\odot}$, the typical distances increase significantly as they become larger than the galactic size, therefore resulting in fewer merging events for a given volume due to the absence the typical Galactic overdensity enhancement~\cite{Pujolas:2021yaw}. A similar effect is observed in the coherent case at heavier black hole masses~\cite{Franciolini:2022htd}. The distances obtained for the stochastic case are significantly larger than in the coherent case and extend to extragalactic regions for heavier PBHs. For clarity, the plot in Fig.~\ref{fig:distancemergers} focuses on the mass range for which the typical merger distances are larger than $10^5\,$kpc. For even larger distances local galactic enhancements of nearby galaxies become relevant and, eventually, the redshift of the signal should also be included for $z \gtrsim 0.1$.

One might expect that the requirement adopted here would lead to a significant loss in the potential reach of the experiment as compared to the true physical signal. However, this is partially balanced by the simple fact that the actual reach of the experiment is now integrated over the full experiment time and not just over the swiping time, see Eq.~\eqref{eq:strain}, therefore partially making up for such a loss. For simplicity, in the following we assume at total observation time of 1\,year. Note, that the following approach allows for a direct utilisation of the axion search data since, due to the small frequency range explored by cavity experiment, changing slightly frequency every few minutes does not alter the magnitude of our estimates.

We now compare the potential reach for both the stochastic and the coherent cases. In this comparison we directly use Eq.~\eqref{eq:strain}, replacing the integration time with the running time of the cavity experiment which, for simplicity, we take to be 1\,year. Since we expect $\mathcal{O}(1)$ mergers to resonate within the cavity at every time, for a fixed BH mass we must use the effective quality factor $Q_{\rm eff}$ in Eq.~\eqref{eq:strain}, analogously to the coherent case. Part of the resonance is suppressed by the presence of multiple sources, potentially resulting in a lower $Q_{\rm eff}$ which is compensated by the presence of multiple detectors.

Regarding the physical source, the difference between the stochastic and the coherent cases is given by the ratio of the distances obtained by adopting Eq.~\eqref{eq:stochcondition} and shown in Fig.~\ref{fig:distancemergers}. Therefore, it is useful to define the potential for a discovery as $S_i=h^i_{\rm physical}/h^i_{\rm reach}$, with $i=$coherent, stochastic. This quantity should be larger than one in order to ensure detection. For simplicity, in the following we focus on the FLASH experiment in Table~\ref{tab:sensitivity} provided that other experiments lead to similar results.

\begin{figure}[!ht]
    \centering
    \includegraphics[width=\linewidth]{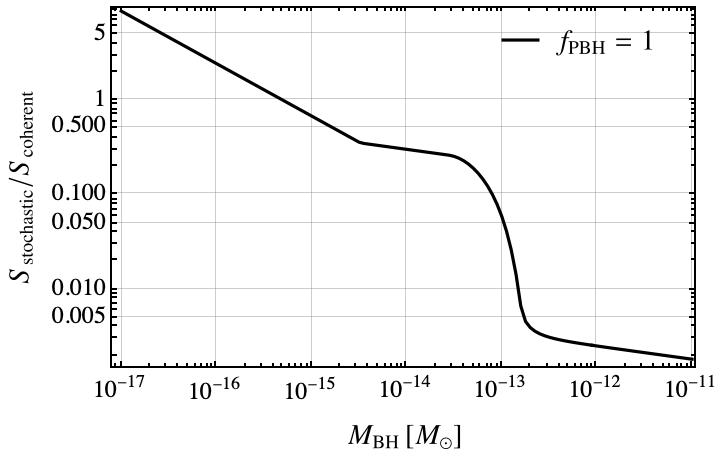}
    \caption{Ratio of potential reach for discovery between stochastic and direct merger detection. See Eq.~\eqref{eq:Ssratio}}.
    \label{fig:ssratio}
\end{figure}

The ratio between the stochastic and coherent potentials for a detection is given by
\begin{equation} 
\label{eq:Ssratio}
	\frac{S_{\rm coherent}}{S_{\rm stochastic}} \simeq \frac{d(1{\rm\,year})}{d(t_{\rm swiping})}\left(\frac{1{\rm\,year}}{t_{\rm eff}^{\rm int}}\right)^{1/4}\,.
\end{equation}
This ratio is shown in Fig.~\ref{fig:ssratio} as a function of the PBH merger masses and for a population with $f_{\rm PBH} = 1$. Similarly to the distance plot comparison in Eq.~\eqref{fig:distancemergers}, there is a dip in the figure coming from the missing galactic enhancement in the PBHs density. At smaller masses the stochastic signal is comparable to the coherent one, therefore suggesting the possibility of a joint search. Note, that the signal distances between the stochastic and coherent case become comparable due to the high density of light PBHs in the galaxy. In this rage of masses, the swiping times are macroscopically large and reach up to $\mathcal{O}(0.1\,{\rm year})$, therefore the condition in Eq.~\eqref{eq:stochcondition} is likely not leading to a clean stochastic background. Likely, a different condition is necessary, possibly leading to potentially larger distances and to smaller values of $Q_{\rm eff}$ due to interference. Consequently, the coherent signal would likely dominate even for light mergers. In any case, in the light mass range window, close to asteroid masses where all of the dark matter could come in the form of primordial black holes, the actual signal is extremely far from the current reach due to the lightness of the primordial masses, as expressed in Fig.~\ref{fig:FLASHreach}.

For higher masses, the stochastic signal gets a suppression due to the large volumes necessary to attain a stochastic background. Therefore, for heavy PBHs it is more feasible to detect a coherent signal from one merger, suggesting the need to develop accurate templates for describing the signal from such events. Our estimate of the stochastic signal is clearly approximate and somewhat optimistic, so we therefore expect these results to remain valid in light of a potential and more accurate analysis.

\section{Conclusion: is there a detection outlook?}
\label{sec:conclusions}

We have discussed the usage of resonant cavities, generally employed as haloscopes to search for the cosmic axion, for the search of gravitational wave signals of astrophysical origin at frequencies (0.1-10)\,GHz. This high-frequency gravitational wave (HFGW) domain spans a significantly broader spectrum compared to what is currently accessible by both ground-based and planned space-based interferometers. Various proposals for these investigations have been advanced in the scientific literature, including one outlined in the FLASH experiment~\cite{Alesini:2023qed}.

If the HFGW signal originates from coalescing primordial black holes (PBHs), cavity experiments can probe the existence of PBH mergers in the mass window $M_{\rm BH} \lesssim 10^{-9}\,M_\odot$. Adopting the setup of the FLASH cavity and considering simultaneous observations of both axions and gravitational waves (GWs) from compact mergers yields the forecast reach as a function of the binary mass in Fig.~\ref{fig:FLASHreach}. Other experiments employing similar setups lead to comparable results.

The signal expected from individual coherent sources in the Galaxy is too faint to be observed with current setups, based on conservative models for the distribution and merging details of the compact objects. This stems from quantifying the loss in the experimental reach due to the actual coherence of the source as discussed in Sec.~\ref{sec:results}. The results do not improve when the collective stochastic signal that originates from the merging of multiple compact objects is considered. In this view, we have given a recipe for the estimate of the stochastic background that differs from previous literature and focuses on the presence of the signal in the cavity at all times, see Sec.~\ref{sec:continuous}. This allows us to predict the region of the PBH masses where the signal is dominated by coherent binary mergers, see Fig.~\ref{fig:ssratio}. Our method relies on the evaluation of the distance that assures to attain at least one event in the cavity at all times, which is obtained from the condition in Eq.~\eqref{eq:stochcondition} and shown in Fig.~\ref{fig:distancemergers} in comparison with the previous literature. Contrary to the coherent case, the distance thus obtained depends on the typical frequency swiping time. Given the vast range of masses in which the detection of a coherent signal would be facilitated over the stochastic background, our findings suggest to push for a broadband search for HFGWs with $\Delta f \sim f$.

Despite this result does not push towards the search of a HFGW signal through resonant cavities, we remind that i) such a search comes at practical no expenses, given that the cavity searches for the cosmic axion are already in place. For this, the detection apparatus should be equipped with an appropriate pick up and acquisition system of the signal generated in cavity modes not excited by the axion, as discussed elsewhere~\cite{Alesini:2023qed}. Moreover, ii) the frequency band covered has not been explored by any dedicated search to date, so that unexpected sources could be present and add up to the GWB as a result of exotic new physics. In view of these considerations, the search for HFGWs is bright.


\begin{acknowledgments}
We thank Diego Blas, Gabriele Franciolini, Gianluca Lamanna, and Francesco Enrico Teofilo for the useful discussions that led to the present work and for reviewing the draft. This research was supported by the M\"unich Institute for Astro-, Particle and BioPhysics (MIAPbP), which is funded by the Deutsche Forschungsgemeinschaft (DFG, German Research Foundation) under Germany's Excellence Strategy - EXC-2094 - 390783311. M.Z.\ and L.V.\ acknowledge support by the National Natural Science Foundation of China (NSFC) through the grant No.\ 12350610240 ``Astrophysical Axion Laboratories'', as well as hospitality by the Istituto Nazionale di Fisica Nucleare (INFN) Frascati National Laboratories and the Galileo Galilei Institute for Theoretical Physics in Firenze (Italy) during the completion of this work. L.V.\ thanks for the hospitality received by the INFN section of Napoli (Italy), the INFN section of Ferrara (Italy), and the University of Texas at Austin (USA) throughout the completion of this work. This publication is based upon work from the COST Actions ``COSMIC WISPers'' (CA21106) and ``Addressing observational tensions in cosmology with systematics and fundamental physics (CosmoVerse)'' (CA21136), both supported by COST (European Cooperation in Science and Technology).
\end{acknowledgments}

\appendix
\section{Detailed resolution of the Einstein-Maxwell equations}
\label{sec:appendixA}

Adapting the derivation in Ref.~\cite{ALPHA:2022rxj} for the search of axions in a resonant cavity, we consider Maxwell's equations in a dielectric medium sourced by a GW strain,
\begin{equation}
    \label{eq:maxwell}
	\begin{split}
	{\bf \nabla}\cdot {\bf D} &= \rho\,,\\
    {\bf \nabla}\times {\bf H} - \partial_t {\bf D} &= {\bf J}\,,\\
    {\bf \nabla}\cdot {\bf B} &= 0\,,\\
    {\bf \nabla}\times {\bf E} + \partial_t {\bf B} &= 0\,.
	\end{split}
\end{equation}
Here, the displacement and the electric fields are related by ${\bf D} = \epsilon {\bf E}$, where $\epsilon$ is the permittivity tensor, while the magnetic flux density and field strength are assumed to coincide, ${\bf B} = {\bf H}$. We solve the set of equations for an infinitely extended cylindrical cavity filled with a plasma, assuming that the axis is aligned along ${\bf \hat{z}}$ where the permittivity $\epsilon_z$ differs from unity.

Under axial symmetry the fields are decomposed as
\begin{equation}
    {\bf B} = {\bf B}_t + B_z {\bf \hat{z}}\,;\quad {\bf E} = {\bf E}_t + \epsilon_z E_z {\bf \hat{z}}\,,
\end{equation}
where the subscripts ``$t$'' and ``$z$'' refer to the orientations transverse and parallel to the z-axis, respectively. Taking the Fourier transform of the fields and sources and assuming that the fields oscillate with frequency $\omega$ gives
\begin{equation}
    \label{eq:Maxwell_intermediate}
    \begin{split}
    & \left({\bf \nabla_t} \!+\! \frac{\partial}{\partial_z}{\bf \hat{z}}\right) \!\times\! \left({\bf B}_t + B_z{\bf \hat{z}}\right) \!=\! - i \omega \left({\bf E}_t + \epsilon_zE_z{\bf \hat{z}}\right) \!+\! \left({\bf J}_t \!+\! J_z{\bf \hat{z}} \right),\\
    & \left({\bf \nabla_t} \!+\! \frac{\partial}{\partial_z}{\bf \hat{z}}\right) \!\times\! \left({\bf E}_t + E_z{\bf \hat{z}}  \right) \!=\! -i \omega \left({\bf B}_t \!+\! \epsilon_zB_z{\bf \hat{z}}\right),
    \end{split}
\end{equation}
where ${\bf \nabla_t}$ is the gradient in the direction transverse to the z-axis. Decomposing Eq.~\eqref{eq:Maxwell_intermediate} into the directions transverse and parallel to ${\bf \hat z}$ yields
\begin{equation}
    \label{eq:decompzt}
    \begin{split}
        {\bf{\hat z}} \cdot {\bf \nabla_t}\times {\bf B_t} &= -i \omega \epsilon_z E_z + J_z\,,\\
        {\bf{\hat z}} \cdot {\bf \nabla_t}\times {\bf E_t} &= i \omega B_z \,,\\
        {\bf{\hat z}}\times \frac{\partial {\bf B_t}}{\partial_z} + {\bf \nabla_t}B_z\times {\bf \hat z} &= -i\omega {\bf E_t} + {\bf J_t}\,,\\
        {\bf{\hat z}}\times \frac{\partial {\bf E_t}}{\partial_z} + {\bf \nabla_t}E_z\times {\bf \hat z} &= i\omega {\bf B_t}\,.
    \end{split}
\end{equation}
The last two expressions in Eq.~\eqref{eq:decompzt} lead to a relation for the transverse fields, once a Fourier transform along the z-axis with momentum $k_z$ is applied, to give
\begin{equation}
\label{eq:etbt}
    \begin{split}
        {\bf E_t} &= \frac{1}{\omega^2 - k_z^2}\left({\bf\nabla_t} \frac{\partial E_z}{\partial z} + i\omega {\bf \nabla_t}B_z\times {\bf \hat{z}} - i \omega {\bf J_t}\right)\,,\\
        {\bf B_t} &= \frac{1}{\omega^2 - k_z^2}\left({\bf\nabla_t} \frac{\partial B_z}{\partial z} - i\omega {\bf \nabla_t}E_z\times {\bf \hat{z}} - \frac{\partial{\bf J_t}}{\partial_z}\right)\,.
    \end{split}
\end{equation}
Contrary to the axion case discussed in Ref.~\cite{ALPHA:2022rxj}, in which the transverse components of the fields are sourced by $z$ components of the current only, in this more general case these components are sourced by the transverse current as well. In fact, in the case of axion ${\bf J}_t=0$ and the $B_z$ component can be neglected. Inserting Eq.~\eqref{eq:etbt} in the first two expressions of Eq.~\eqref{eq:decompzt} gives
\begin{equation}
    \label{eq:eoms}
    \begin{split}
    &\frac{\omega^2}{\omega^2-k_z^2}{\bf \nabla_t}^2 E_z + \omega^2 \epsilon_z E_z + i\omega J_z = \frac{i\omega}{\omega^2-k_z^2}{\bf\hat{z}}\cdot{\bf \nabla_t}\times \frac{\partial {\bf J}_t}{\partial z}\,,\\
    &\frac{\omega^2}{\omega^2-k_z^2}{\bf \nabla_t}^2 B_z - \omega^2 B_z = \frac{\omega^2}{\omega^2-k_z^2}{\bf \hat{z}}\cdot{\bf \nabla_t}\times {\bf J}_t\,.
    \end{split}
\end{equation}
Note, that the negative sign in front of the second term of $B_z$ equation can lead to tachyonic instability, which are avoided when the dissipation in the cavity walls are considered.

We now specialize these results to a resonant cavity by expanding the electric field in modes of pulsation $\omega_n$ labeled by an integer $n$, as
\begin{equation}
    E_z = \sum_n e_n\, E_n({\bf x})\,,    
\end{equation}
where $e_n$ is the coefficient of the expansion and the eigenmodes satisfy~\cite{ALPHA:2022rxj}\footnote{Contrary to Ref.~\cite{ALPHA:2022rxj}, we normalize the volume integral without the volume factor on the right-hand side.}
\begin{equation}
\begin{split}
    &{\bf \nabla_t}^2 E_n= \epsilon_z'\left(k_z^2-\omega_n^2 \right)\,E_n\,,\\
    &\int{\rm d}^3{\bf x}\, E_n E_m=\delta_{nm}\,.
    \end{split}
\end{equation}
Here, the permeability has been decomposed as $\epsilon_z = \epsilon_z' - i \epsilon_z''$ following standard notation~\cite{Landau1961ElectrodynamicsOC}. Since the induced current has frequency $\omega_n$,  $\epsilon'' = \omega_n/(\omega\, Q )$, therefore leading, from Eq.~\eqref{eq:eoms}, to
\begin{equation}
    \label{eq:eomdecomp}
    \omega^2 \epsilon_z' \frac{\omega^2-\omega_n^2}{\omega^2 - k_z^2}\,e_n - i\frac{\omega\,\omega_n}{Q}\,e_n = -i\omega\int {\rm d}^3 {\bf x}\,E_n\,\Tilde{J}_t\,,
\end{equation}
where we defined
\begin{equation}
    \Tilde{ J}_t = J_z - \frac{1}{\omega^2-k_z^2}{\bf \hat z}\cdot {\bf\nabla_t}\times \frac{\partial \bf J_t}{\partial z}\,.
\end{equation}
Solving for $e_n$ we finally obtain
\begin{equation}
    \label{eq:Ecomponent}
    e_n = \frac{-i\omega}{(\omega^2 -\omega_p^2) \frac{\omega^2-\omega_n^2}{\omega^2 - k_z^2} - i\frac{\omega\,\omega_n}{Q}}\,\int {\rm d}^3 {\bf x}\, {E}_n  \Tilde{ J}_t\,.
\end{equation}
Following Eq.~(20) in Ref.~\cite{Berlin:2021txa}, we decompose the current as
\begin{equation}
	\Tilde{ J}_t = B_0 \omega^2 V^{1/3} \sum_a h_a(\omega) \hat{\bf j}_a\,,
\end{equation}
so that
\begin{equation}
    \label{eq:Ecomponent1}
    e_n = \frac{ -i B_0 \omega^3 V^{1/3}}{(\omega^2 -\omega_p^2) \frac{\omega^2-\omega_n^2}{\omega^2 - k_z^2} - i\frac{\omega\,\omega_n}{Q}}\, \sum_A h_A(\omega)\,\left(\int {\rm d}^3 {\bf x}\, {E}_n \hat{\bf j}_A \right) \,.
\end{equation}
This leads to the signal
\begin{equation}
    \label{eq:Ecomponent2}
    e_n = -i \omega\,B_0\,V^{5/6}\,\mathcal{T}(\omega)\, \sum_A h_A(\omega)\,\eta_A  \,,
\end{equation}
where we introduced the cavity-GW coupling coefficient
\begin{equation}
	\eta_A = \frac{1}{V^{1/2}}\int {\rm d}^3 {\bf x}\, {E}_n \hat{\bf j}_A \,.
\end{equation}
In Eq.~\eqref{eq:Ecomponent2}, the function controlling the resonance is given by
\begin{equation}
    \mathcal{T}(\omega) = \left[\left(1 -\frac{\omega_p^2}{\omega^2}\right) \frac{\omega^2-\omega_n^2}{\omega^2 - k_z^2} - \frac{i}{Q}\frac{\omega_n}{\omega}\right]^{-1}\,.
\end{equation}

The energy stored in the $n$-th cavity mode is
\begin{equation}
	U = \int {\rm d}f{\rm d}f'\,\langle e_n(\omega)\,e_n(\omega')\rangle\,,
\end{equation}
or, using the expression above,
\begin{eqnarray}
    \label{eq:energystored1}
	U &=& \int {\rm d}f{\rm d}f'\,
	\omega(\omega')\,B_0^2 \,V^{5/3}\,\mathcal{T}(\omega)\mathcal{T}^*(\omega')\nonumber\\
    &&\times \sum_{AA'} \langle h_A(\omega)h_{A'}(\omega') \rangle \,\eta_A\,\eta_{A'}\,.
\end{eqnarray}
For a stochastic signal, we write
\begin{equation}
	\langle h_A(\omega)h_{A'}(\omega') \rangle = \frac{3H_0^2}{32\pi^3}\delta_{AA'}\delta(f-f')\,f^{-3}\,\Omega_{\rm GW}(\omega)\,,
\end{equation}
so that when $k_z = 0$ we find
\begin{equation}
    \label{eq:energystored}
	U = \int \frac{{\rm d}\omega}{2\pi} \,\frac{(4\pi) \,(2\pi)^3\,\omega\,B_0^2\,V^{5/3}\,\eta^2\,\Omega_{\rm GW}(\omega)}{\omega^2\left[\left(1-\frac{\omega_p^2}{\omega^2}\right) \,\left(1-\frac{\omega_n^2}{\omega^2}\right)\right]^2 + \left(\frac{\omega_n}{Q}\right)^2}\,\frac{3H_0^2}{32\pi^3}\,,
\end{equation}
where an extra $(4\pi)$ comes from the angular integration and an extra $(2\pi)^3$ from converting $f$ into $\omega$. We have also introduced the coupling
\begin{equation}
    \label{eq:defineeta}
    \eta = \left(\sum_{A} \,\eta_A^2\right)^{1/2} \approx 0.14\,,
\end{equation}
where the last expression corresponds to the cavity mode TM$_{012}$. Inserting the expression for $\Omega_{\rm GW}$ in Eq.~\eqref{eq:OmegaGW} into Eq.~\eqref{eq:energystored} gives
\begin{equation}
    \label{eq:energystored2}
	U = \frac{1}{2}B_0^2V^{5/3}\eta^2\int_0^{+\infty}\!\!{\rm d}\omega \frac{\omega^3\,h_c^2}{\omega^2\left(1\!-\!\frac{\omega_p^2}{\omega^2}\right)^2\!\left(1\!-\!\frac{\omega_n^2}{\omega^2}\right)^2 \!+\! \left(\frac{\omega_n}{Q}\right)^2}\,.
\end{equation}
We first consider a coherent source with $N_{\rm cycle}$ cycles that can be observed inside the cavity, so that the strain is modeled as~\cite{Hong:2014vua, OHare:2017yze}
\begin{equation}
    \label{eq:strainhc}
    h_c^2 = \frac{h_0^2}{4 \pi N_{\rm cycle}} \frac{\omega_n^2}{(\omega - \omega_n)^2 + \omega_n^2/(2 N_{\rm cycle})^2}\,.
\end{equation}
To estimate the signal power in the cavity $P_{\rm sig} = \omega U/Q$, we first evaluate the integral in Eq.~\eqref{eq:energystored2} around $\omega \approx \omega_n$ in the case $\omega_p = 0$. This leads to the expression for the power in the signal as
\begin{equation}
    \label{eq:signal}
    P_{\rm sig} \approx \frac{1}{4}B_0^2V^{5/3}\eta^2 \omega_n^3\,h_0^2\,Q_{\rm eff}\,,\quad\hbox{for $\omega_p  = 0$}\,,
\end{equation}
where $Q_{\rm eff} = \min(N_{\rm cycle}, Q)$. This result coincides with the findings in previous literature, apart for an overall numerical factor $\mathcal{O}(1)$~\cite{Berlin:2021txa}.

On the other hand, the GWB is estimated by assuming a constant $h_c$ in the integrand of Eq.~\eqref{eq:energystored2}, in place of Eq.~\eqref{eq:strainhc}. Note, that in the case where $\omega_n \approx \omega_p$, which is the condition under which the ALPHA cavity experiment is going to operate, there is an additional enhancement by approximately a factor $Q$ that is relevant when detecting the GWB,
\begin{equation}
    \label{eq:signalALPHA}
    P_{\rm sig} \approx \sqrt{\frac{\pi}{6}}\,B_0^2\,V^{5/3}\,\eta^2\,\omega_n^3\,h_c^2\,Q\,,\quad\hbox{for $\omega_p \approx \omega_n$}\,.
\end{equation}

We now estimate the strain sensitivity in the cavity. To estimate the signal-to-noise ratio (SNR), we account for the instrument noise which is characterized by the power
\begin{equation}
    \label{eq:noise}
    P_{\rm noise} =  T_{\rm sys}\,\sqrt{\frac{\Delta f}{\Delta t}}\,,
\end{equation}
where $T_{\rm sys}$ is the coldest effective temperature of the instrumentation and noise amplifier. The ratio between Eqs.~\eqref{eq:signal} and~\eqref{eq:noise} expresses the SNR as
\begin{equation}
    {\rm SNR} = P_{\rm sig}/P_{\rm noise}\,.
\end{equation}
Setting ${\rm SNR} =1$ leads to the estimate for the reach of the GW strain in the resonant cavity as in Eq.~\eqref{eq:strain} when focusing on the setup where $\omega_p = 0$.

\bibliographystyle{apsrev4-1}
\bibliography{HFGW.bib}

\end{document}